\documentclass[prb,twocolumn, superscriptaddress,preprintnumbers,amsmath,amssymb showpacs,]{revtex4}

\usepackage{graphicx}
\usepackage{latexsym}
\usepackage{dcolumn}
\usepackage{bm}
\usepackage{float}
\usepackage{graphicx,here}

\begin{document}

\title{Direct Evidence of Metallic Bands in a Monolayer Boron Sheet}

\author{Baojie Feng}
\email{bjfeng@issp.u-tokyo.ac.jp}
\affiliation{Institute for Solid State Physics, The University of Tokyo, Kashiwa, Chiba 277-8581, Japan}
\author{Jin Zhang}
\affiliation{Institute of Physics, Chinese Academy of Sciences, Beijing 100190, China}
\author{Ro-Ya Liu}
\affiliation{Institute for Solid State Physics, The University of Tokyo, Kashiwa, Chiba 277-8581, Japan}
\author{Takushi Iimori}
\affiliation{Institute for Solid State Physics, The University of Tokyo, Kashiwa, Chiba 277-8581, Japan}
\author{Chao Lian}
\affiliation{Institute of Physics, Chinese Academy of Sciences, Beijing 100190, China}
\author{Hui Li}
\affiliation{Institute of Physics, Chinese Academy of Sciences, Beijing 100190, China}
\author{Lan Chen}
\affiliation{Institute of Physics, Chinese Academy of Sciences, Beijing 100190, China}
\author{Kehui Wu}
\affiliation{Institute of Physics, Chinese Academy of Sciences, Beijing 100190, China}
\affiliation{Collaborative Innovation Center of Quantum Matter, Beijing 100871, China}
\author{Sheng Meng}
\email{smeng@iphy.ac.cn}
\affiliation{Institute of Physics, Chinese Academy of Sciences, Beijing 100190, China}
\affiliation{Collaborative Innovation Center of Quantum Matter, Beijing 100871, China}
\author{Fumio Komori}
\affiliation{Institute for Solid State Physics, The University of Tokyo, Kashiwa, Chiba 277-8581, Japan}
\author{Iwao Matsuda}
\email{imatsuda@issp.u-tokyo.ac.jp}
\affiliation{Institute for Solid State Physics, The University of Tokyo, Kashiwa, Chiba 277-8581, Japan}

\date{\today}


\begin{abstract}
The search for metallic boron allotropes has attracted great attention in the past decades and recent theoretical works predict the existence of metallicity in monolayer boron. Here, we synthesize the $\beta_{12}$-sheet monolayer boron on a Ag(111) surface and confirm the presence of metallic boron-derived bands using angle-resolved photoemission spectroscopy. The Fermi surface is composed of one electron pocket at the $\rm\overline{S}$ point and a pair of hole pockets near the $\rm\overline{X}$ point, which is supported by the first-principles calculations. The metallic boron allotrope in $\beta_{12}$-sheet opens novel physics and chemistry in material science.
\end{abstract}

\pacs{ }

\maketitle

Boron, the fifth element in the periodic table, has been known to be the lightest element substance that forms interatomic covalent bonds. Since the bonding states in the bulk boron ranges from two-center to multi-center bonds, boron forms varieties of allotropes that show rich physical and chemical properties\cite{Woods1994,Albert2009,Albert2013}. Although boron has been discovered for more than two centuries, its crystalline form was synthesized only half a century ago\cite{Sands1957,Hughes1963,McCarty1958,Talley1960}. Up to now, all the allotropes of boron have been found to be semiconducting with a large band gap\cite{Albert2009,Albert2013}. Recently, the discovery of graphene\cite{Neto2009} has triggered great interest in the search for elemental monolayer materials and theoretical investigations on the related materials have spurred experimental findings of silicene\cite{Vogt2012,Feng2012,Tao2015}, germanene\cite{Li2014,Davila2014,Derivaz2015}, and phosphorene\cite{Wang2015}. Likewise, monolayer boron has also been explored theoretically \cite{Tang2007,Tang2009,Penev2012,Ozdogan2010,LiuH2013,Liu2013,Zhang2015,Wu2012} and, intriguingly, it has been predicted to be metallic which sharply contrasts the semiconducting nature of the bulk allotropes. Very recently, experimental synthesis of monolayer boron was realized by directly evaporating pure boron on the Ag(111) surface\cite{Feng2016,Mannix2015}. Although the metallic property of the boron monolayer on Ag(111) was implied by scanning tunneling spectroscopy (STS), it was not conclusive as the Ag(111) substrate is highly conductive\cite{Feng2016,Mannix2015}.

In this Letter, we report the results of band mapping by angle-resolved photoemission spectroscopy (ARPES), which directly reveal metallic bands in the top boron layer in this system. By comparing low-energy electron diffraction (LEED) patterns with the STM observation and first-principles calculations\cite{Feng2016}, the boron layer is identified as the so-called $\beta_{12}$-sheet that contains ordered pattern of hexagon holes\cite{Tang2007,Tang2009}. The photoemission Fermi surface includes one electron pocket centered at the $\rm\overline{S}$ point and a pair of electron pockets near the $\rm\overline{X}$ point in the two-dimensional Brillouin zone (2D BZ). The band dispersion curves are in agreement with those calculated by the first-principles calculations on the $\beta_{12}$-sheet structure model. Finding of the metallic bands of monolayer boron is expected to promote further investigations of novel properties such as the superconductivity predicted theoretically in the present $\beta_{12}$-sheet structure model\cite{Penev2016}. Moreover, a selective nitridation of the boron monolayer is expected to synthesize insulating boron nitrides that forms insulator-metal junction in the same single layer, enabling the development of all-boron nano-devices.

\begin{figure}[b]
\includegraphics[width=8cm]{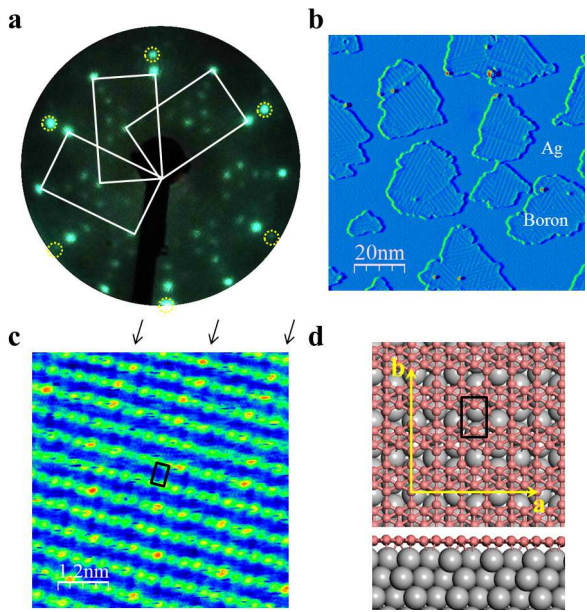}
\caption{(a) A LEED pattern of monolayer boron sheet grown on Ag(111) taken with 65 eV. The spots of Ag(111)-1$\times$1 are indicated by the yellow circles, and the reciprocal lattices of boron are illustrated by the white rectangles. Note that there are three equivalent orientations of boron domains rotated by 60$^{\circ}$. (b) Derivative STM image of the boron sheet. The striped patterns are the moir\'{e} patterns arising from the mismatch of the boron sheet and Ag(111). (c) High resolution STM image (0.9 V) showing the atomic resolution of the boron sheet. The black arrows indicate the striped patterns observed in (b). (d) Structure model of the monolayer boron sheet on Ag(111): top and side view. The black rectangles in (c) and (d) correspond to the unit cell of the monolayer boron sheet. }
\end{figure}

The sample was prepared by directly evaporating pure boron (99.9999$\rm\%$) onto a Ag(111) substrate. Single crystal Ag(111) was cleaned by repeated sputtering and annealing cycles, and the cleanness of the surface was confirmed by sharp LEED patterns and ARPES measurements of the Shockley surface state. During the growth of boron, the substrate temperature was controlled at about 600 K. The purity of the as-prepared sample was confirmed by \textit{in-situ} X-ray photoelectron spectroscopy (XPS) using an Al K$\alpha$  X-ray source\cite{Supp}. The ARPES measurements were performed at 120 K with a Scienta SES100 analyzer and helium discharge lamp (He I$\alpha$ light). The Fermi level was calibrated by measurements of the spectral Fermi-edge on the clean Ag(111) surface. STM experiments were performed at 77 K in another system to examine the surface structure under the same preparation conditions.

First-principles calculations, using density functional theory (DFT), were performed with the Vienna Ab initio Simulation Package (VASP)\cite{Krasse1996}. The projector augmented-waves method\cite{Blochl1994} and Perdew-Burke-Ernzerhof (PBE) exchange-correlation\cite{Krasse1996} were used. The plane-wave cutoff energy was set to be 400 eV and the vacuum space was set to be larger than 15 $\rm\AA$. As the PBE function usually overestimates the chemical bond length, the lattice constant of Ag(111) used in the calculations was 3\% larger than the experimental value. The boron $\beta_{12}$-sheet model (Fig.1(d)) contains a boron monolayer on a four-layer 3$\rm \times 3 \sqrt{3}$ Ag(111) surface. The 2D BZ was sampled using the Monkhorst-Pack scheme\cite{Monkhorstl1976}. A $k$-point mesh 5$\rm\times$5$\rm\times$1 was used for the structural optimization, while a 6$\rm\times$6$\rm\times$1 mesh was adopted in the self-consistent calculations. Using the conjugate gradient method, the positions of the atoms are optimized until the residual force on each atom is less than 0.05 eV/$\rm\AA$. The calculated band structure was further processed by the orbital-selective band unfolding technique\cite{Medeiros2014,Medeiros2015}. Due to the high surface-sensitivity of ARPES, the calculated band structure of boron/Ag(111) is projected onto the boron layer for comparison with the ARPES spectra.

Figure 1(a) shows a LEED pattern of monolayer boron on Ag(111). Besides the 1$\rm\times$1 spots of Ag(111), as indicated by the yellow dotted circles, one can recognize additional spots which originate from the boron sheets. The monolayer boron takes a rectangular lattice with three equivalent orientations, as indicated by the white rectangles. The other darker spots arise from the moir$\rm\Acute{e}$ patterns, induced by the lattice mismatch between the boron sheets and the Ag(111) substrate. From the LEED pattern, the lattice constants of the boron sheets are 5.0 $\rm\AA$ and 2.9 $\rm\AA$ along the $\rm [1\overline{1}0]$ and $\rm [\overline{1}\overline{1}2]$ directions of Ag(111), respectively. The results are consistent with the STM image that shows parallel striped moir$\rm\Acute{e}$ patterns in three equivalent orientations, as shown in Fig.1(b). The high-resolution STM image in Fig.1(c) confirms a rectangular lattice of the boron monolayer. According to the previous STM work\cite{Feng2016}, this rectangular boron structure corresponds to the $\beta_{12}$-sheet\cite{LiuH2013,Zhang2015}, as shown in Fig.1(d). It should be noted that there is another monolayer boron structure reported recently\cite{Mannix2015}, which also exhibits a rectangular lattice with the same lattice constants in the STM images. However, its orientation is 30$^\circ$ rotated with respect to the structure in the present research, which means that the monolayer boron we synthesized is different from the one reported in Ref.\cite{Mannix2015}.

\begin{figure*}[htb]
\includegraphics[width=14cm]{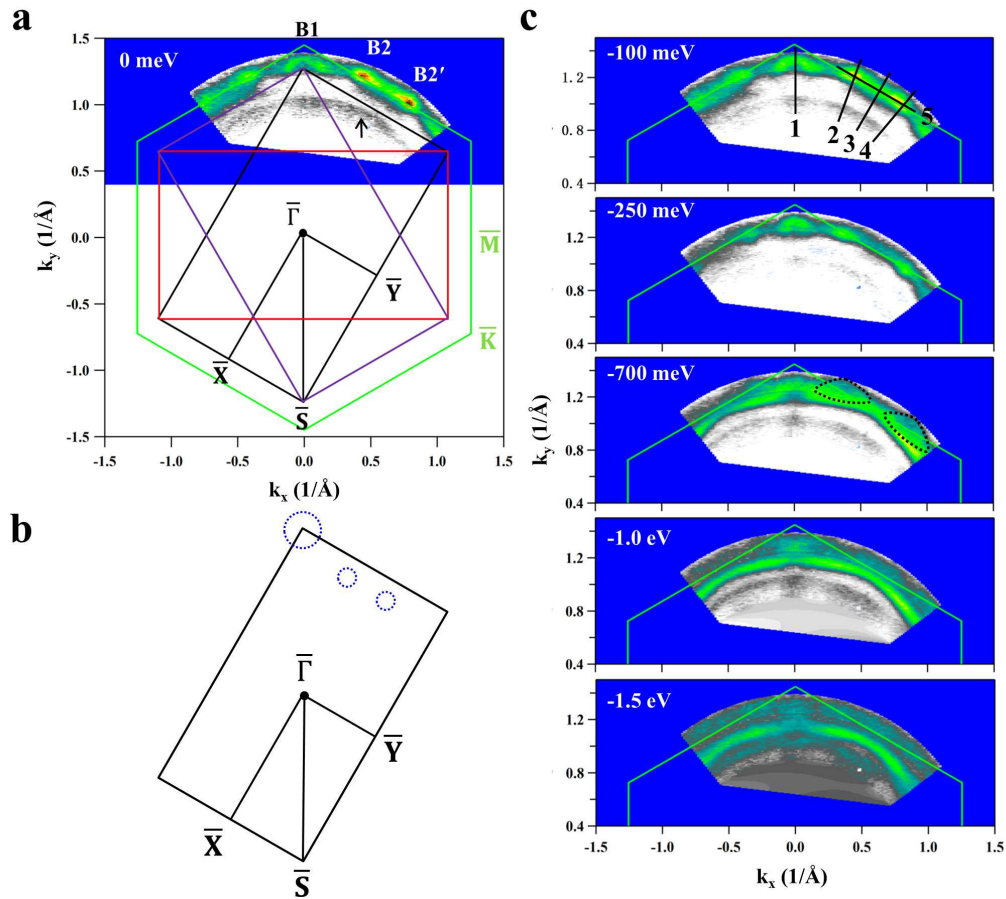}
\caption{ Constant energy contours (CEC's) of the monolayer boron on Ag(111), measured by photoemission spectroscopy. (a) At Fermi surface ($\rm E_{F}-E$ = 0 meV). The black solid arrow indicates the $sp$ band of Ag(111). The green hexagon indicates the BZ of Ag(111); the black, purple, and red rectangles indicate the BZ of monolayer boron for three equivalent directions. (b) Schematic drawing of the Fermi surfaces with the Brillouin zone of the boron sheet. (c) CECs of monolayer boron at different binding energies:$\rm E_{F}-E$ = -100 meV, -250 meV, -700 meV, -1.0 eV and -1.5 eV.}
\label{fig2}
\end{figure*}

\begin{figure*}[htb]
\includegraphics[width=18cm]{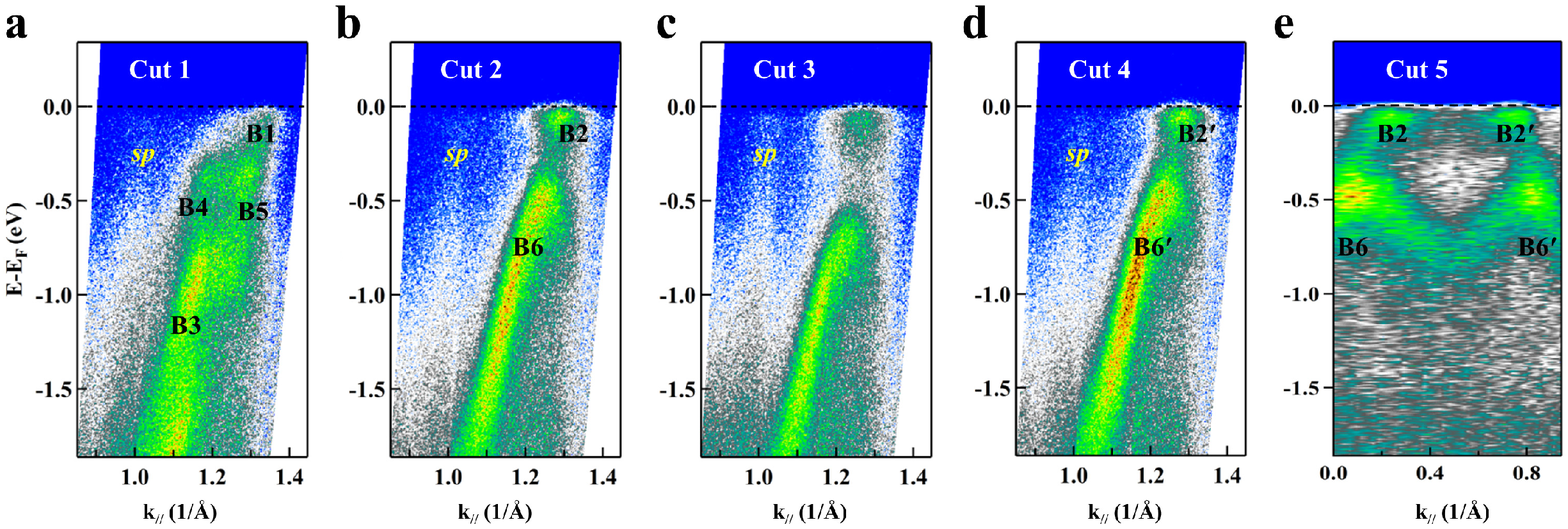}
\caption{(a-e) ARPES intensity plots along cut 1 to cut 5 indicated in Fig. 2(c). The bands of the boron sheet are labeled by B1 to B6. In cut 1 to cut 4, one can also observe the bulk $sp$ bands of Ag(111) simultaneously with the bands of boron, as marked by ``\textit{sp}".}
\label{fig3}
\end{figure*}

The photoemission Fermi surfaces of the boron monolayer on Ag(111) are shown in Figure 2(a), which appears six-fold symmetric owing to the existence of three equivalent domains on the substrate. For clarity, the Fermi surface data were superimposed with the 2D BZ of both Ag(111) and the boron sheets. The green hexagon in Fig.2(a) corresponds to the 2D BZ of Ag(111); the black, yellow and red rectangles correspond to those of the boron sheets in three equivalent directions. Three spectral features, indicated as B1, B2, and B2', appear in the photoemission Fermi surface mapping after preparation of the boron monolayer on Ag(111). This result clearly indicates that the boron layer, the $\beta_{12}$-sheet, is metallic. As shown in Fig.2(b), these Fermi surfaces are located at the $\rm\overline{S}$ point (B1) and near the $\rm\overline{X}$ point (B2 and B2'). The total carrier concentration ($\rm n_{B}$) of the monolayer boron can be evaluated from the area of the Fermi pockets and it was found to be $\rm n_{B} \sim 3 \times 10^{13} cm^{-2}$. In the experimental Fermi surface map of Fig.2(a), remnant signals of the Ag(111) $sp$ bands are also observed as indicated by the black arrow. Spectral changes of the Ag surface state and bulk bands with increasing coverage of boron sheets are described in the supplementary information\cite{Supp}. Figure 2(c) shows photoemission constant energy contours (CECs) of the boron monolayer on Ag(111), taken at different binding energies, $E-E_{F}$. Comparing CECs at the Fermi level ($\rm E_{F}$), $E-E_{F}$=-100 meV, and $E-E_{F}$=-250 meV, one can recognize that the areas of the B1, B2, and B2' bands become smaller with increasing binding energy. This allows us to assign that these Fermi surfaces are electron pockets. On the other hand, new spectral features, as illustrated by the black dashed lines, emerge in CEC's at deeper binding energies ($E-E_{F}$=-700 meV, -1.0 eV, -1.5 eV), indicating the existence of other electronic bands in these energy region.

\begin{figure*}[htb]
\includegraphics[width=14cm]{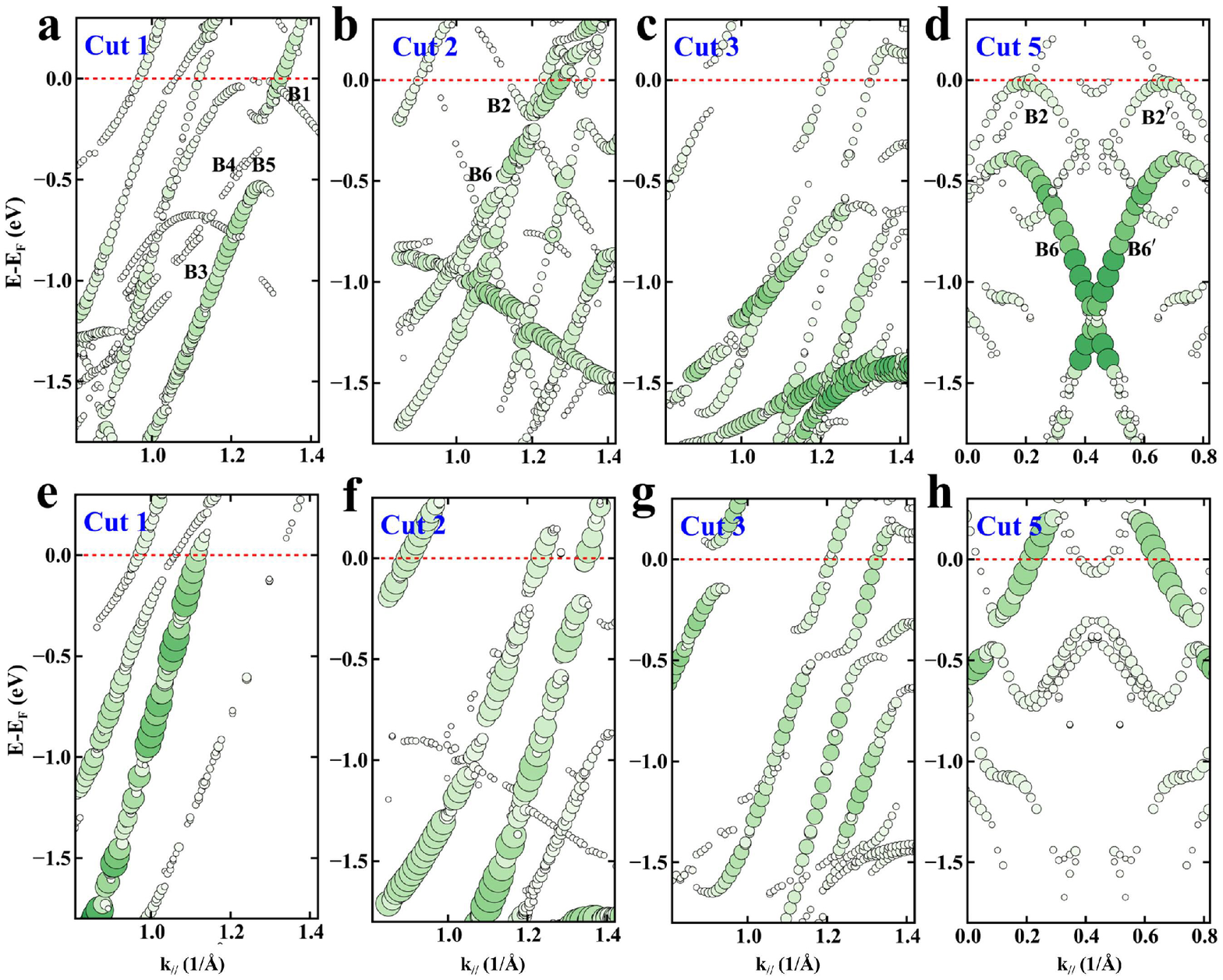}
\caption{(a-d) Bands of boron/Ag(111) projected on the boron sheet along cuts 1, 2, 3 and 5, respectively. (e-h) Bands of boron/Ag(111) projected on Ag(111) along cuts 1, 2, 3 and 5, respectively. The size of the dots denotes the spectral weight. The black dashed lines are guides for the eye to compare with the bands observed in the ARPES results. }
\label{fig4}
\end{figure*}

To clarify electronic structure of these bands in detail, Figure 3 shows the band dispersion plots along several line cuts illustrated in Fig.2(c). Along a Cut 1 of Fig.3(a), an electronic band B1 is observed near $\rm E_{F}$, as confirmed in Fig.2. At higher binding energies, one can find a number of band-dispersion curves that are composed of, at least, three bands, B4, B5, and B6. Along Cut 2 and Cut 4, one can observe a pair of electronic bands (B2 and B2'), corresponding to the two electron pockets near the $\rm\overline{X}$ point in Fig.2. Sharply dispersing bands B6 and B6' can be identified at higher binding energies. In the figures, remnant signals of the Ag(111) bulk $\textit{sp}$ band can be identified. Since the bands of the Ag(111) substrate are well-separated from the electron pockets of the boron sheet and maintain its original dispersion curve, any hybridization of the boron sheet with Ag(111) seems to be weak, which is in accordance with the previous theoretical and experimental results\cite{Liu2013,Zhang2015,Feng2016}. As a result, the bands, B1-B6, of the boron monolayer are essentially preserved on Ag(111).

To further understand the band structure of the monolayer boron sheet, first-principles calculations\cite{Monkhorstl1976,Krasse1996,Blochl1994} were performed together with a band-unfolding analysis\cite{Medeiros2014,Medeiros2015}. Figures 4(a-d) and (e-h) show the calculated band structures of the monolayer boron on Ag(111) along different Cuts (1, 2, 3 and 5) projected onto the boron layer and the Ag(111) substrate, respectively. In Fig.4(a-d), one can trace dispersion curves of the B1, B2(B2'), B3 and B6(B6') bands, as illustrated by the black dashed curves. Features of the B4 and B5 bands can also be found in Fig.4(a). These band features of the boron layers in Fig.4(a-c) are distinctive from those of the Ag(111) substrate in Fig.4(e-g). Thus, one can confirm that the spectral features of B1-B6 correspond to the boron layer. On the other hand, along Cut 5, there is apparent spectral weight near B2(B2') from Ag(111), as shown in Fig.4 (h). This indicates that the band of B2(B2') along this direction mainly originates from the Ag(111) beneath the boron layer. Indeed, there always exists interaction between overlayer and substrate at any surface. Our calculation results identify the momentum regions or the corresponding wave functions that are involved in such interaction in the present boron/Ag(111) system. As shown in Fig.4, there are some bands in the calculation that were not observed in the photoemission diagram (Fig.3). This is likely due to the matrix element effect in the photoemission process and the spectral appearance can depend on various factors such as energy and polarization of the incident photon.

In summary, the present results unequivocally confirm the metallic character of the monolayer boron sheets, which will pave the way for new physics and chemistry in material science. Specifically, metallicity of the boron sheet opens an avenue toward two-dimensional superconductivity as predicted theoretically for the present $\beta_{12}$-sheet structure model\cite{Penev2016}. The large electron-phonon coupling in monolayer boron may even lead to a high superconducting transition temperature. We also suggest that regulated or selective nitridation of the boron monolayer could lead to patterned formation of insulating boron nitride useful for building insulator-metal junctions in the same single layer. Such systems could promote technical innovations of novel all-boron nano-devices.

\begin{acknowledgements}
We thank Prof. T.-C. Chiang for helpful discussions. We also thank Prof. X.J. Zhou for providing the Igor macro to analyze the ARPES data. This work was supported by the Ministry of Education, Culture, Sports, Science and Technology of Japan (Photon and Quantum Basic Research Coordinated Development Program), a Japan Society for the Promotion of Science grant-in-aid for specially promoting research ($\rm\#$23000008) and for Scientific Research (B) ($\rm\#$26287061), and by JST, ACT-C.
B.F and J.Z. contributed equally to this work.
\end{acknowledgements}

\end{document}